\documentclass [12pt] {article}

\begin{document}

\title{%
Testing Dispersion Relations of Quantum $\kappa$--Poincar\'e
Algebra on Cosmological Ground}
\author{ J.\ Kowalski--Glikman\thanks{e-mail
address jurekk@ift.uni.wroc.pl}\\ Institute for Theoretical
Physics\\ University of Wroc\l{}aw\\ Pl.\ Maxa Borna 9\\
Pl--50-204 Wroc\l{}aw, Poland} \date{October 26, 2000} \maketitle

\begin{abstract}
Following the procedure proposed recently by Martin and
Brandenberger  we investigate  the spectrum of the cosmological
perturbations in the case when the ``trans--Plackian'' dispersion
relations are motivated by the quantum $\kappa$-Poincar\'e
algebra. We find that depending on the choice of initial
conditions of the perturbations, the spectrum  either differs from
the flat one for for instantaneous Minkowski vacuum  or, in the
case of initial conditions minimizing energy density,  leads to
the observed scale-invariant Harrison--Zel'dovich spectrum in the
Friedmann epoch.
\end{abstract}
\vspace{12pt}

\clearpage

In the recent years a growing mass of evidence appeared,
indicating that on short (trans-Planckian) scale the usual
space-time symmetries are drastically modified, eg., by existence
of a fundamental length scale (see e.g.  \cite{dofrro,gara}). Some
time ago  Lukierski,  Nowicki,  Ruegg, and  Tolstoy
\cite{lunoruto} using formalism of quantum algebras derived a
quantum deformation of the Poincar\'e algebra, which is a high
energy extension of the standard low-energy Poincar\'e algebra.
This algebra, called $\kappa$-Poincar\'e algebra includes the
parameter $\kappa$ of dimension of (inverse) length, usually
identified with the  Planck length. The algebra  (in the so-called
bicrossproduct basis)
 \cite{maru,luruto})
 takes the form:

\begin{eqnarray} \label{1}
 [M_{\mu\nu},M_{\rho\tau}]
=&& \displaystyle i\left(\eta_{\mu\tau}M_{\nu\rho} -
\eta_{\nu\rho}M_{\nu\tau} +\eta_{\nu\rho}M_{\mu\tau}
  -\  \eta_{\nu\tau}M_{\mu\rho} \right),
\cr \cr
 [M_{i},P_{j}] =&&  i\epsilon_{ijk}P_{k}\, ,
\quad [M_{i},P_{0}]= iP_{i}\, , \cr \cr \displaystyle
  [N_{i}, P_{j}] = && -i\delta_{ij}
 \left( {\kappa\over 2} \left(
 1 -e^{2{P_{0}/ \kappa}}
\right) + {1\over 2\kappa} \vec{P}\, ^{ 2}\, \right) + \ {i\over
\kappa} P_{i}P_{j} , \cr \cr
 \left[N_{i},P_{0}\right] = && iP_{i}\, ,
\cr \cr \displaystyle
  [P_{\mu},P_{\nu}] = && 0\, ,
\label{1.1a}
\end{eqnarray}
where $P_\mu =(P_i, P_0)$ are space and time components of
four-momentum, and $M_{\mu\nu}$ are modified Lorentz generators
with rotations $M_k = \frac12\epsilon_{ijk}M_{ij} $, and boosts
$N_i = M_{0i}$. It can be readily found that the first Casimir of
this algebra, defining the mass-shell is of the
form\footnote{There exists another realization of this quantum
algebra, with slightly different co-product, for whose the first
Casimir is of the form $\mathcal{C}^{bcp'}_1 = (\vec{P})^2 e^{+
P_0/\kappa} -
 \left(2\kappa\sinh\left(\frac{P_0}{2\kappa}\right)\right)^2$ and
 the dispersion relation $(k)^2 e^{+ \omega/\kappa} -
 \left(2\kappa\sinh\left(\frac{\omega}{2\kappa}\right)\right)^2=0$.
 This case differs from the cases considered in this papers by the
 fact that there is a cut-off for three-momentum:
 $k\rightarrow\kappa$ when $\omega\rightarrow\infty$. For this
 reason we will consider this case in a separate paper.}
\begin{equation}\label{2}
 \mathcal{C}^{bcp}_1 = (\vec{P})^2 e^{- P_0/\kappa} -
 \left(2\kappa\sinh\left(\frac{P_0}{2\kappa}\right)\right)^2.
\end{equation}

Equations (\ref{1}, \ref{2}), with additional structures of
quantum algebra (coproducts and antipodes) form the algebraic
framework for field theoretical constructions. In the first step
in this construction one should identify the deformed
$\kappa$-Minkowski space. This can be done by taking the set of
generators dual to the $\kappa$-Poincar\'e algebra, $(\hat x^\mu,
\Lambda^{\rho\sigma})$, which form the $\kappa$-Poincar\'e group
and dividing it by its  Lorentz subgroup (observe that in algebra
(\ref{1}) the Lorentz generators are not deformed.) This has been
done in the papers \cite{maru, szak}. As a results one obtains
that the deformed $\kappa$-Minkowski space must be necessarily a
non-commuting space:
\begin{equation}\label{a}
 \left[\hat x^\mu, \hat x^\nu\right]= \frac{i}{\kappa} \left(\delta^\mu_0
 \hat x^\nu -\delta^\nu_0
 \hat x^\mu\right)
\end{equation}

Now the question arises as to how one can define a quantum theory
on such a space. Such a construction has been presented recently
in \cite{koslukmas}. This construction consists of three major
steps. First one $\kappa$-deforms the classical local field theory
on standard Minkowski space obtaining as a result a {\em local}
$\kappa$-deformed field theory on $\kappa$-deformed, {\em
non-commutative} Minkowski space. Then one observes that momenta
in (\ref{1}) are commutative and uses an appropriately defined
$\kappa$-deformed Fourier transform to obtain a $\kappa$ deformed
field theory on commutative momentum space. In the last step one
makes use of the standard inverse Fourier transform to get a {\em
non-local}, $\kappa$-deformed field theory on {\em standard}
Minkowski space. This non-locality exhibits itself in
non-polynomial structure of the Casimir operator (\ref{2}), but
the fact that the Minkowski space is now standard makes it
possible to use the standard realization of momentum operators as
derivatives over commuting Minkowski space positions, $P_\mu = i
\partial/\partial x^\mu$.

One can now turn to the free massless scalar field (see
\cite{koslukmas}). The invariant wave operator on
$\kappa$-deformed Minkowski space $\frac{\partial}{\partial \hat
x_\mu}\frac{\partial}{\partial \hat x^\mu}$ can be expressed, in
the momentum space as $\mathcal{C}^{bcp}_1 \left(1 -
\frac{\mathcal{C}^{bcp}_1}{4\kappa^2}\right)$. We see that the
spectrum of the modified massless wave operator contains the
deformed massless mode $\mathcal{C}^{bcp}_1 \phi =0$ and the
tachyon $\left(\mathcal{C}^{bcp}_1 - 4\kappa^2\right) \phi =0$. In
this paper we will consider only the first branch, leaving the
possible physical meaning of the second one to future
investigations, and thus we are left with a field theory whose
dynamics is governed by the Casimir operator, (\ref{2}).

This  Casimir operator  leads to the following dispersion relation
\begin{equation}\label{4}
  (k)^2 e^{- \omega/\kappa} -
 \left(2\kappa\sinh\left(\frac{\omega}{2\kappa}\right)\right)^2=0.
\end{equation}
It should be noted that the wave equation leading to the above
dispersion relation is non-local in time. Following the discussion
in \cite{luruza} we solve this equation for $\omega^2$ so that the
resulting dispersion relation corresponds to an operator of second
order in time derivatives, and non-local in space:
\begin{equation}\label{5}
  \omega^2 = \left[\kappa\log\left(1+\frac k\kappa\right)\right]^2.
\end{equation}
Below we will use this expression to define a theory of modified
dynamics of fluctuations in inflationary universe. It should be
noted that except of the second Newton's law there is no clear
principle saying which relation is physically more fundamental.
Therefore, strictly speaking one cannot say that the theory
described by (\ref{5}) and investigated below is derived from
$\kappa$-Poincar\'e algebra, though it is not excluded that one
could derive this theory from the first principles using some
procedure different from that presented in \cite{koslukmas}.

It is worth pausing for a moment to show that the dispersion
relation (\ref{5}) indeed leads to modified wave equation
covariant under $\kappa$-Poincar\'e symmetry. For the scalar field
$\phi$ this equation reads
\begin{equation}\label{5a}
 P_0^2 \phi = \left[\kappa\log\left(1+\frac
{|\vec{P}|}\kappa\right)\right]^2 \phi .
\end{equation}
It is clear that this equation is invariant under
three-dimensional rotations generated by $M_i$. Let us consider
generalized boosts parameterized by infinitesimal parameter
 $\varepsilon$ resulting from eq.~(\ref{1}).
\begin{equation}\label{5b}
\delta_\varepsilon P_0 = \left[\varepsilon^i N_i,P_0 \right] =  i
\varepsilon^i P_i
\end{equation}
\begin{equation}\label{5c}
\delta_\varepsilon P_j = \left[\varepsilon^i N_i,P_j \right] =
-i\varepsilon_{j}
 \left( {\kappa\over 2} \left(
 1 -e^{2{P_{0}/ \kappa}}
\right) + {1\over 2\kappa} \vec{P}\, ^{ 2}\, \right) + \ {i\over
\kappa} \varepsilon^i P_{i}P_{j}
\end{equation}
One can check by direct computation that eq.~(\ref{5a}) is
covariant under transformations (\ref{5b}), (\ref{5c})
\footnote{In the course of this calculation one uses
eq.~(\ref{5a}) twice: first to express $P_0 \delta P_0$ as
$\kappa\log\left(1+\frac {|\vec{P}|}\kappa\right)\delta P_0$ and
second to rewrite the  $ e^{2{P_{0}/ \kappa}}$ factor in terms of
$P_i$. }.
\newline

Since the parameter $\kappa$ is assumed to be of order of Planck
length, for a long time the $\kappa$-Poincar\'e algebra remained
 a nice mathematical construction without any direct, testable
 physical applications and consequences (see however
 \cite{tests}). Recently however some areas have been identified,
 where the behavior of matter at very high, ``trans--Planckian''
 frequencies has a direct, low energy consequences. One of these
 areas is the black hole physics. The question whether modified
 high frequency behavior has an effect on Hawking radiation and
 Hawking--Beckenstein entropy formula has
 been considered (in the context of sonic black holes) by Unruh
 \cite{unruh},  Corley and Jacobson, and others \cite{jac} (for recent review
 see \cite{jacrev}). These authors proposed a rather ad hoc
 dispersion relations, to wit
\begin{equation}\label{6}
\omega = \kappa \tanh^{1/p}
\left[\left(\frac{k}{\kappa}\right)^{p}\right], \quad \mbox{
(Unruh) }
\end{equation}
and
\begin{equation}\label{7}
\omega^2 = k^2 - \frac{k^4}{\kappa^2}, \quad \mbox{
(Corley--Jacobson)},
\end{equation}
where we again denoted by $\kappa$ the characteristic length.
\newline

In the recent papers Niemeyer \cite{Niem} and Martin and
Branderberger \cite{mb} considered another setting in which there
is a direct relation between trans--Planckian physics and low
energy phenomenology. Namely, they considered quantum fluctuations
during inflationary stage, giving rise to large scale formation in
the Friedmann universe. These fluctuations are enormously red
shifted in the course of inflation, and thus the wavelengths of
the modes that correspond to the present large scale structure
were, at the initial time, well in the realm of the
trans--Planckian physics. It follows that it is justified to ask,
to what extend the modified dispersion relations have their
imprints on the present day large scale structure of the universe.
Since we know that in order to agree with observations, the
spectrum of fluctuations should be (almost) flat, this setting is
perfect to test validity of modified dispersion relations, derived
or motivated by other means. One should stress at this point that
the relation (\ref{4}) we are to investigate in this paper are
motivated by considerations, having root in some fundamental
physics (in the case at hands, from quantum groups).

To set the stage let us recall some fundamental facts concerning
early cosmology and large scale structure formation. We consider
spatially flat Friedmann universe with the conformally flat line
element
\begin{equation}\label{8}
 ds^2 = a^2(\eta)\left( -d\eta^2 + \sum_{i=1}^3 dx_i^2\right).
\end{equation}
The conformal time $\eta$ is related to the cosmic time $t$ by
relation $dt = a(\eta)d\eta$; in particular the De--Sitter
universe corresponds to $\eta= H^{-1}e^{-Ht}$, $a(\eta) =
l_H/\eta$.

To compute power spectra of observable quantities in the standard
case (i.e., with unmodified dispersion relations) one considers
equation of evolution of modes of Fourier components of massless
scalar field $\mu_n$
\begin{equation}\label{9}
 \mu''_n + \left( n^2 - \frac{a''}{a}\right)\mu_n=0,
\end{equation}
where prime denotes derivative with respect to conformal time
$\eta$.  When the wavelength of fluctuations is much longer than
the characteristic cosmological scale, the Hubble radius,
$\lambda(\eta) \equiv a(\eta) 2\pi/n \gg l_H$, so that the first
term in the parenthesis is small compared to $a''/a$,  the
solution of this equation is $\mu_n(\eta) = C_n a(\eta)$. One can
then calculate the resulting spectra, to obtain
\begin{equation}\label{10}
 n^3 P = n^3 |C_n|^2
\end{equation}
The observed, flat, scale invariant spectrum corresponds to the
case when $C_n \sim n^{-3/2}$.

This behaviour can be readily seen from the explicit general
solution of eq.~(\ref{9}). One has in this case
\begin{equation}\label{10a}
  \mu_n(\eta) = \alpha_n \left(1 + \frac{i}{n\eta}\right) e^{in\eta} +
  \beta_n \left(1 - \frac{i}{n\eta}\right) e^{-in\eta}.
\end{equation}
For large $\eta$ (corresponding to early times) the second term in
parentheses is small and the fluctuation behaves as a wave with
slowly changing amplitude; starting from $\eta_H = 2\pi/n$,
corresponding to the moment when fluctuation crosses the horizon,
 there are no oscillations and the wave gets frozen. Thus to find
 the spectrum we have to compute
\begin{equation}\label{cn}
 |C_n| = \frac{1}{a(\eta_H)}\,
 \left|\mu_n\left(\eta_H\right)\right|=
 \frac{2\pi}{l_H} \frac{1}{n} \left|\mu_n\left(\eta_H\right)\right|.
\end{equation}
We see therefore that in order to lead to flat spectrum, the
coefficients $\alpha_n$, $\beta_n$ should behave as $\sim 1/\sqrt
n$.
\newline

Equation (\ref{9}) corresponds to the standard dispersion relation
$$\omega^2 = k^2 = \frac{n^2}{a^2}.$$ In the case of modified
dispersion relation $\omega^2 = \Upsilon^2(k)$ one should replace
$n^2$ with
\begin{equation}\label{11}
 n^2_{mod} = a^2(\eta)\Upsilon^2(k) =
 a^2(\eta)\Upsilon^2\left(\frac{n}{a(\eta)}\right).
\end{equation}
Therefore the equation we are to consider is
\begin{equation}\label{12}
\mu''_n + \left( a^2(\eta)\Upsilon^2\left(\frac{n}{a(\eta)}\right)
- \frac{a''}{a}\right)\mu_n=0.
\end{equation}
This equation should be appended by initial conditions at for
$\mu_n$ and $\mu'_n$. In this paper we will consider two natural
sets of initial conditions: the ``minimal energy condition'' of
\cite{mb},
\begin{equation}\label{13}
 \mu_n(\eta_i) =
 \sqrt{\frac{a(\eta_i)}2}\Upsilon^{-1/2}\left(\frac{n}{2\pi a(\eta_i)}\right),
\end{equation}
\begin{equation}\label{14}
 \mu'_n(\eta_i) =\pm
 i\sqrt{\frac{1}{2a(\eta_i)}}\Upsilon^{1/2}\left(\frac{n}{2\pi a(\eta_i)}\right);
\end{equation}
and the instantaneous Minkowski vacuum conditions, as in the case
of standard dispersion relation
\begin{equation}\label{14a}
  \mu_n(\eta_i) =
 \frac1{\sqrt{2n}}, \quad \mu'_n(\eta_i) =\pm i
 \sqrt{\frac n{2}}.
\end{equation}
Observe that in the case of standard dispersion relation,
$\Upsilon^2=n^2$ these initial condition are identical.

 Our goal would be therefore to solve eq.~(\ref{12}) with initial
 conditions (\ref{13}, \ref{14}) or (\ref{14a}), and by making use of expression
 (\ref{10}) to find the power spectrum of cosmological
 perturbation. We will consider two regimes: (I) when
 modifications of dispersion relations are relevant and $\Upsilon$
 term dominates in eq.~(\ref{12}); (II) when the perturbation is
 large as compared to the scale defined by $\kappa$, in which case we have to do
 with standard dispersion relation and eq.~(\ref{10a}) holds. With
 this equation we can compute modulus of the coefficient  $C_n$.

Let us turn to the dispersion relation (\ref{5}) which is valid in
regime (I)
\begin{equation}\label{17}
\Upsilon(k) = \kappa \log \left(1+ \frac{k}{\kappa}\right).
\end{equation}
 In this regime, assuming $a(\eta) =
l_H/\eta$ eq.~(\ref{12}) takes the form
\begin{equation}\label{19}
 \frac{d^2\mu^{(I)}_n}{d\eta^2} + \left[\frac{4\pi^2\epsilon^2}{\eta^2}
 \log^2\left(1+\frac{n \eta}{2\pi\epsilon}\right) - \frac2{\eta^2}
 \right]\mu^{(I)}_n=0,
\end{equation}
where we introduced the parameter $\epsilon=(\kappa l_H)$ which is
a ratio of two relevant length scales $l_H$, the size of the
cosmological horizon and $1/\kappa$. If we assume that $1/\kappa$
is of order of Planck length, then $\epsilon$ is equal to the size
of the horizon expressed in Planck unit. The numerical value of
this parameter depends therefore on details of the dynamics of
inflation. Following \cite{mb} we assume that $\epsilon\sim
10^{6}$.

To solve  equation (\ref{19}) we make use of the fact that for
equation
 $\mu'' + W^2(\eta)\mu$ the solution  is of the approximate form
 $$\mu= \frac{1}{\sqrt{2 W}}\,\exp\left(\pm i\int W(\eta) d\eta\right).$$ This
 solution is valid in the adiabatic regime, where
\begin{equation}\label{19a}
 \frac12\left(\frac{W''}{W^3} - \frac32\frac{W'{}^2}{W^4}\right)
 \ll 1.
\end{equation}
In our case we get the  condition $${\frac{-3\,{n^2}\,{{\eta }^2}
+
     4\,n\,\pi \,\epsilon \,\eta \,
      \log (1 + {\frac{n\,\eta }{2\,\pi \,\epsilon }}) +
     {{\left( 2\,\pi \,\epsilon  + n\,\eta  \right) }^2}\,
      {{\log (1 + {\frac{n\,\eta }{2\,\pi \,\epsilon }})}^
        2}}{16\,{{\pi }^2}\,{{\epsilon }^2}\,
     {{\left( 2\,\pi \,\epsilon  + n\,\eta  \right) }^2}\,
     {{\log (1 + {\frac{n\,\eta }{2\,\pi \,\epsilon }})}^
       4}}}\ll 1.$$
In regime (I) the fluctuations are described therefore by $$
\mu^{(I)}_n = A^{(I)}_n
\sqrt{\frac{\eta}{4\pi\epsilon\log\left(1+\frac{n
\eta}{2\pi{\epsilon}}\right)}}\exp\left[-2\pi i\epsilon
Li_2\left(-\frac{n\eta}{2\pi\epsilon}\right)\right]$$
\begin{equation}\label{20}+B^{(I)}_n
\sqrt{\frac{\eta}{4\pi\epsilon\log\left(1+\frac{n
\eta}{2\pi\epsilon}\right)}}\exp\left[+2\pi i\epsilon
Li_2\left(-\frac{n\eta}{2\pi\epsilon}\right)\right],
\end{equation}
where $Li_2(x) \equiv \int dx \log(1-x)/x$ is the polylogarithm
function.

Let us now turn to initial conditions (\ref{13}, \ref{14}). At the
time $\eta = \eta_i$ we find $$ \mu^{(I)}_n(\eta_i) = A^{(I)}_n
\sqrt{\frac{\eta_i}{4\pi\epsilon\log\left(1+\frac{n
\eta_i}{2\pi\epsilon}\right)}}\exp\left[-2\pi i\epsilon
Li_2\left(-\frac{n\eta_i}{2\pi\epsilon}\right)\right]$$
\begin{equation}\label{21}+B^{(I)}_n
\sqrt{\frac{\eta_i}{4\pi\epsilon\log\left(1+\frac{n
\eta_i}{2\pi\epsilon}\right)}}\exp\left[+2\pi i\epsilon
Li_2\left(-\frac{n\eta_i}{2\pi\epsilon}\right)\right]
=\sqrt{\frac{\eta_i}{4\pi\epsilon \log \left(1+
 \frac{n \eta_i}{2\pi\epsilon}\right)}}
\end{equation}and
$$\frac{1}{i} \mu^{(I)}_n{}'(\eta_i) = A^{(I)}_n
\sqrt{\frac{2\pi\epsilon\log\left(1+\frac{n
\eta_i}{2\pi\epsilon}\right)}{2\eta_i}}\exp\left[-2\pi i\epsilon
Li_2\left(-\frac{n\eta_i}{2\pi\epsilon}\right)\right]$$
\begin{equation}\label{22}-B^{(I)}_n
\sqrt{\frac{2\pi\epsilon\log\left(1+\frac{n
\eta_i}{2\pi\epsilon}\right)}{2\eta_i}}\exp\left[+2\pi i\epsilon
Li_2\left(-\frac{n\eta_i}{2\pi\epsilon}\right)\right]
=\sqrt{\frac{2\pi\epsilon \log \left(1+
 \frac{n \eta_i}{2\pi\epsilon}\right)}{2\eta_i}},
 \end{equation}
Which leads to
\begin{equation}\label{23}
A^{(I)}_n = \exp\left[2\pi i\epsilon
Li_2\left(-\frac{n\eta_i}{2\pi\epsilon}\right)\right], \quad
B^{(I)}_n=0,
\end{equation}
and
\begin{equation}\label{24}
\mu^{(I)}_n = \sqrt{\frac{\eta}{4\pi\epsilon\log\left(1+\frac{n
\eta}{2\pi\epsilon}\right)}}\exp\left[2\pi i\epsilon
Li_2\left(-\frac{n\eta_i}{2\pi\epsilon}\right)-2\pi i\epsilon
Li_2\left(-\frac{n\eta}{2\pi\epsilon}\right)\right].
\end{equation}
In the case of linear standard vacuum initial conditions,
(\ref{14a}) we similarly obtain
\begin{equation}\label{25}
A_n^{(I)\, st} =
\left(\frac1{\sqrt{8n}}\sqrt{\frac{2\pi\epsilon\log\left(1+\frac{n
\eta_i}{2\pi\epsilon}\right)}{2\eta_i}} +
\sqrt{\frac{n}{8}}\sqrt{\frac{\eta_i}{4\pi\epsilon\log\left(1+\frac{n
\eta_i}{2\pi\epsilon}\right)}}\right)e^{2\pi i\epsilon
Li_2\left(-\frac{n\eta_i}{2\pi\epsilon}\right)}
\end{equation}
\begin{equation}\label{26}
B_n^{(I)\, st} =
\left(\frac1{\sqrt{8n}}\sqrt{\frac{2\pi\epsilon\log\left(1+\frac{n
\eta_i}{2\pi\epsilon}\right)}{2\eta_i}} -\sqrt{\frac{n}{8}}
\sqrt{\frac{\eta_i}{4\pi\epsilon\log\left(1+\frac{n
\eta_i}{2\pi\epsilon}\right)}}\right)e^{-2\pi i\epsilon
Li_2\left(-\frac{n\eta_i}{2\pi\epsilon}\right)}
\end{equation}

Now we turn to regime (II), and we have to match above solution
with $\mu_n^{(II)}$ given by (\ref{10a})  at $\eta = \eta_1$.
Before doing that, let us pause for a moment to find the
characteristic value of the conformal time $\eta_1$. It
corresponds to the moment when the wavelength of the perturbation
equals the characteristic length $\kappa^{-1}$. Thus
\begin{equation}\label{30}
 \eta_1 = \frac{2\pi\epsilon}{n}.
\end{equation}
It must be checked if at this value of conformal time we are still
in the region of validity of adiabatic approximation, i.e., if
condition (\ref{19a}) is satisfied. One finds the condition
$${\frac{-3 + 4\,{{\log (2)}^2} + \log (4)}
   {32\,{{\pi }^2}\,{{\epsilon }^2}\,{{\log (2)}^4}}} \ll 1$$
   which indeed holds for any $\epsilon \sim 10^{-1}$ and larger.
   On the other hand we must also satisfy the condition $$4\pi^2
   \epsilon^2 \log(2) \gg 2$$ so that we can safely assume that
   fluctuations do not feel the cosmic expansion and neglecting the factor $(-2/\eta^2)$ in
   eq.~(\ref{19}) was indeed justified. This condition
   is well satisfied for $\epsilon$ slightly larger than $1$. We
   see therefore that our approximate solution holds at the
   matching point $\eta = \eta_1$. Physically this means that for such values of the parameter
   $\epsilon$, the fluctuation
    relaxes sufficiently slowly from the regime ruled by the trans-Planckian
   physics and at the same time reaches the regime described by standard physics
   well before the fluctuation length becomes comparable with the
   Hubble size.
   This analysis of the
   adiabaticity conditions is in perfect agreement with the results
   of papers \cite{Niem}, \cite{mb}.
\newline

   Let us thus match the solution (\ref{24}) with the solution
   (\ref{10a}). In view of eq.~(\ref{cn}) we will be mainly
   interested in the $n$ dependence of the coefficients. Computing
   $\mu^{(I)}_n(\eta_1)$ and its derivative and comparing with
   appropriate expressions obtained from (\ref{cn}) we easily find
\begin{equation}\label{31}
\alpha_n \sim \frac1{\sqrt n}\, e^{2\pi i\epsilon
Li_2\left(-\frac{n\eta_i}{2\pi\epsilon}\right)}, \quad \beta_n
\sim \frac1{\sqrt n}\, e^{2\pi i\epsilon
Li_2\left(-\frac{n\eta_i}{2\pi\epsilon}\right)},
\end{equation}
both multiplied by complicated numerical ($n$-independent)
coefficients. This means that $$|C_n| \sim n^{-3/2}$$ and we
recover the flat spectrum of the fluctuations.

For instantaneous Minkowski vacuum initial conditions (\ref{14a}),
the final result changes. One finds $$\alpha_n, \beta_n \sim const
+ \frac1n,$$ and thus one cannot obtain the scale invariant
spectrum. This corresponds to the result  of Martin and
Brandenberger \cite{mb} who also concluded that these standard
initial conditions do not lead to the correct spectrum of
fluctuations.

This result is not very surprising. The  initial
  conditions (\ref{14a}) are natural in the
case of standard dispersion relation. However, it would be hardly
understandable if they remain valid in the case of modified
nonlinear dispersion. The choice of the minimal energy conditions
of Martin and Brandenberger (\ref{13}, \ref{14}), on the other
hand, is justified by powerful physical principle that the
fluctuations leading to large scale structure originate from
 vacuum fluctuations with minimal possible energy.
\newline

 This result is the first test passed by $\kappa$ physics.
This is important because till now $\kappa$-Poincar\'e symmetry
was nothing but a nice algebraic construction. I hope  that the
result reported in this paper would encourages one to look for new
grounds where predictions of modified dispersion relations might
be possibly checked (for a very recent attempt along this line,
see \cite{a-cp}.)

Some open problems remain, of course. Probably the most important
one is what is the source of ``trans-Plackian frequencies
reservoir''. In other words one should understand the initial
conditions (\ref{13}, \ref{14}) from the point of view of
fundamental $\kappa$-quantum field theory. It is also interesting
to see imprints of $\kappa$-physics on early stages of inflation,
especially in the pictures where the inflation starts immediately
after Planck era.
\newline

{\bf Acknowledgement}. I would like to thank Professor Jerzy
Lukierski for enlightening discussions concerning $\kappa$
physics, and to the anonymous referee, whose comments made it
possible to improve this paper greatly.

\end{document}